\documentclass[twocolumn,aps, prl,showpacs]{revtex4}

\usepackage{graphics}% Include figure files
\usepackage{graphicx}% Include figure files
\usepackage{dcolumn}% Align table columns on decimal point
\usepackage{bm}% bold math
\usepackage{amsmath,amssymb}

\newcommand{\de}{\partial}
\newcommand{\sech}[1]{\textrm{sech}\left(  #1\right)}
\newcommand{\eq}[2]{\begin{equation} \label{#1} #2 \end{equation}}

\newcommand{\etal}{{\em et al.}}

\begin{document}

\title{High-energy, shock-front assisted resonant radiation in the normal dispersion regime}
\author{Thomas Roger$^{1}$, Mohammed F. Saleh$^{2}$, Samudra Roy$^{2}$, Fabio Biancalana$^{1,2}$, Chunyong Li$^{1}$ and Daniele Faccio$^{1}$}
\affiliation{$^{1}$School of Engineering and Physical Sciences, Heriot-Watt University, EH14 4AS Edinburgh, UK \\
$^{2}$Max Planck Institute for the Science of Light, G\"{u}nther-Scharowsky str. 1, 91058 Erlangen, Germany}
\date{\today}

\begin{abstract}
We present a simple yet effective theory that predicts the existence of resonant radiation bands in the deep normal group velocity dispersion region of a medium, even in absence of a zero-group velocity dispersion point. This radiation is evident when the medium is pumped with high-energy ultrashort pulses, and it is driven by the interplay between the Kerr and the shock terms in the NLSE. Accurate experiments performed in bulk silica fully support the theoretical phase-matching condition found by our theory. 
\end{abstract}
\pacs{42.65.Wi, 42.65.Tg, 42.65.Jx, 42.65.-k, 42.65.Sf}
\maketitle

%\paragraph{Introduction ---}

In nonlinear optics, when a  temporal soliton propagates in a medium with additional perturbations (such as Raman effect, higher-order dispersions, external potentials, etc.), it may emit, under special circumstances, a kind of radiation known as {\em resonant radiation} (RR), when the momentum of the soliton matches the propagation constant of the waveguide itself \cite{akhmediev,fission,biancalana}. This {\em phase-matching} occurs typically only for specific frequencies, that can be well separated from the central frequency of the soliton, allowing a powerful energy transfer from the soliton to the radiation. \\
There has been a recent surge of interest in the phenomenon leading to the formation of resonant radiation in fibers and bulk media. In recent experiments, ultraviolet resonant radiation emitted by blue-shifting solitons in hollow-core photonic crystal fibers has been demonstrated \cite{biancalanablue}. Moreover, resonant radiation has been demonstrated to be responsible, amongst other mechanisms, for supercontinuum generation in optical fibers, and in particular in photonic crystal fibers \cite{fission,akhmediev,biancalana,dudley}. Resonant radiation can also be observed in micron-size integrated waveguides \cite{colman}, in silicon-on-insulator waveguides \cite{dmitry} and even in waveguide arrays, where it is emitted by spatial solitons \cite{tranwaveguides}. In hydrodynamics, shallow water solitons can also emit linear waves \cite{karpman}. This universality has also led to fresh interpretations in the basic processes leading to the formation and control of resonant radiation in optics \cite{erkintalo,NRR}.\\
In previous works on the subject, it was argued that in order to generate dispersive resonant radiation, one does not strictly need a soliton, since the radiation can be emitted even when pumping in the normal dispersion regime in the presence of a zero-group velocity dispersion (GVD) point sufficiently close to the pump wavelength \cite{roy,webb}. The phase-matching condition between the pulse and the radiation is essentially identical to the soliton case \cite{webb}. All these works assume the impossibility of creating dispersive waves in the absence of higher-order dispersion (HOD) terms, since these are traditionally the sources of the perturbation needed to excite the radiation modes \cite{akhmediev,fission,biancalana}.\\
In this Letter, we show that when operating in the normal dispersion regime of a medium it is not necessary to have higher-order dispersion or a zero-GVD point to produce a powerful source of resonant radiation that is well-separated from the pump. We have found, rather surprisingly, that only the Kerr effect is needed and that the phase-matching condition is influenced strongly by a previously overlooked effect, namely shock-induced dispersion. We derive a new resonant radiation phase-matching condition and we support our theory with numerical simulations. Moreover, we demonstrate experimentally the existence of the new shock-front induced resonant radiation, by using pulses propagating in normal GVD in bulk silica.\\
\emph{Phase-matching condition for resonant radiation in normal GVD --}
Propagation of intense ultrashort pulses in a nonlinear medium is described by the following GNLSE:
\eq{NLSE}{i\de_{z}A+\hat{D}(i\de_{t})A+\left[1+\hat{S}(i\de_{t})\right]\gamma|A|^{2}A=0,} where $z$ and $t$ are the longitudinal spatial and temporal coordinates respectively, $A(z,t)$ is the electric field envelope,  $\hat{D}(i\de_{t})\equiv\sum_{m=2}^{\infty}\beta_{m}(i\de_{t})^{m}/m!$ is the dispersion operator, $\hat{S}(i\de_{t})\equiv(i/\omega_{0})\de_{t}$ is the shock operator, $\beta_{m}$ is the $m$-th order dispersion coefficient, $\gamma$ is the nonlinear coefficient of the medium, and $\omega_{0}$ is the central input pulse frequency.\\
Resonant radiation will be emitted at frequencies for which the propagation constant of the medium matches the input pulse momentum, whether the pulse is a soliton or not. This condition can be easily found by equating the total phase of the pump ($\phi_{0}$) and the total phase of the radiation ($\phi_{\rm R}$) after a common delay $t\equiv z/v_{\rm g}(\omega_{0})$, with $v_{\rm g}\equiv\beta_{1}(\omega_{0})^{-1}$, where $\beta(\omega)$ is the propagation constant of the medium:
\begin{eqnarray}
\phi_{0}&=&\left\{\beta(\omega_{0})-\omega_{0}/v_{\rm g}(\omega_{0})+\gamma P\left[1+\frac{\omega_{\rm R}-\omega_{0}}{\omega_{0}}\right]\right\}z\label{phase1}\\
\phi_{\rm R}&=&\left\{\beta(\omega_{\rm R})-\omega_{\rm R}/v_{\rm g}(\omega_{0})\right\}z\label{phase2}
\end{eqnarray} where $P$ is the power associated to the maximum intensity in the pulse evolution, and is therefore $z$-dependent (unlike the soliton case).
The third term inside the curly brackets in Eq. (\ref{phase1}) is due to the combined action of the self-phase modulation and the shock operator. The latter is traditionally neglected in the literature since it is supposed to modify the resonant condition only slightly, but in this Letter we show that it plays a crucial role in the absence of a zero-GVD point in the vicinity of the pump frequency $\omega_{0}$.
The phase matching condition for the radiation emission, namely $\phi_{0}=\phi_{\rm R}$, results in the following equation for the frequency detuning $\Delta\omega\equiv\omega_{\rm R}-\omega_{0}$ between pulse and radiation, obtained by expanding in Taylor series the propagation constant of the radiation around $\omega_{0}$:
\eq{phasematching}{\sum_{m=2}^{\infty}\frac{1}{m!}\beta_{m}\Delta\omega^{m}=\gamma P\left[1+\frac{\Delta\omega}{\omega_{0}}\right]}
If only the second order dispersion coefficient ($\beta_{2}$) is taken into account on the left hand side of Eq. (\ref{phasematching}), one obtains the following two resonant radiation frequencies in physical units:
\eq{fullsolution}{\Delta\omega_{\pm}=\frac{\gamma P\pm\sqrt{(\gamma P)^{2}+2\beta_{2}\gamma P\omega_{0}^{2}}}{\omega_{0}\beta_{2}}}
The first solution, $\Delta\omega_{+}$, is always positive, and thus the resonant radiation associated to this solution is always blue-shifted with respect to the pump frequency. The second solution, $\Delta\omega_{-}$, is always negative and close to zero, so that the radiation associated to it is always red-shifted and very close to the pump frequency and for this reason may be rather difficult to observe experimentally.
An important and surprising point is that it is not necessary to have a zero-GVD point in the dispersion in order for pulses to emit resonant radiation. Only the Kerr effect, when combined with the pulse propagating in the normal GVD regime of the medium is sufficient to produce such radiation, which will appear as two identical peaks, spectrally symmetric with respect to the pump. Note that self phase modulation (SPM), when combined with the normal GVD, induces the formation of lateral peaks known as `optical wave breaking' \cite{owb}, which were noted both experimentally and numerically, and are attributed to four-wave mixing (FWM) \cite{owb,agrawal2}. However, in this Letter we give a very different explanation in terms of phase-matching to a new type of resonant radiation. The possibility of the generation and control of such radiation has been overlooked in previous studies and recent works on the subject \cite{webb,roy}.

\begin{figure}
\includegraphics[width=8cm]{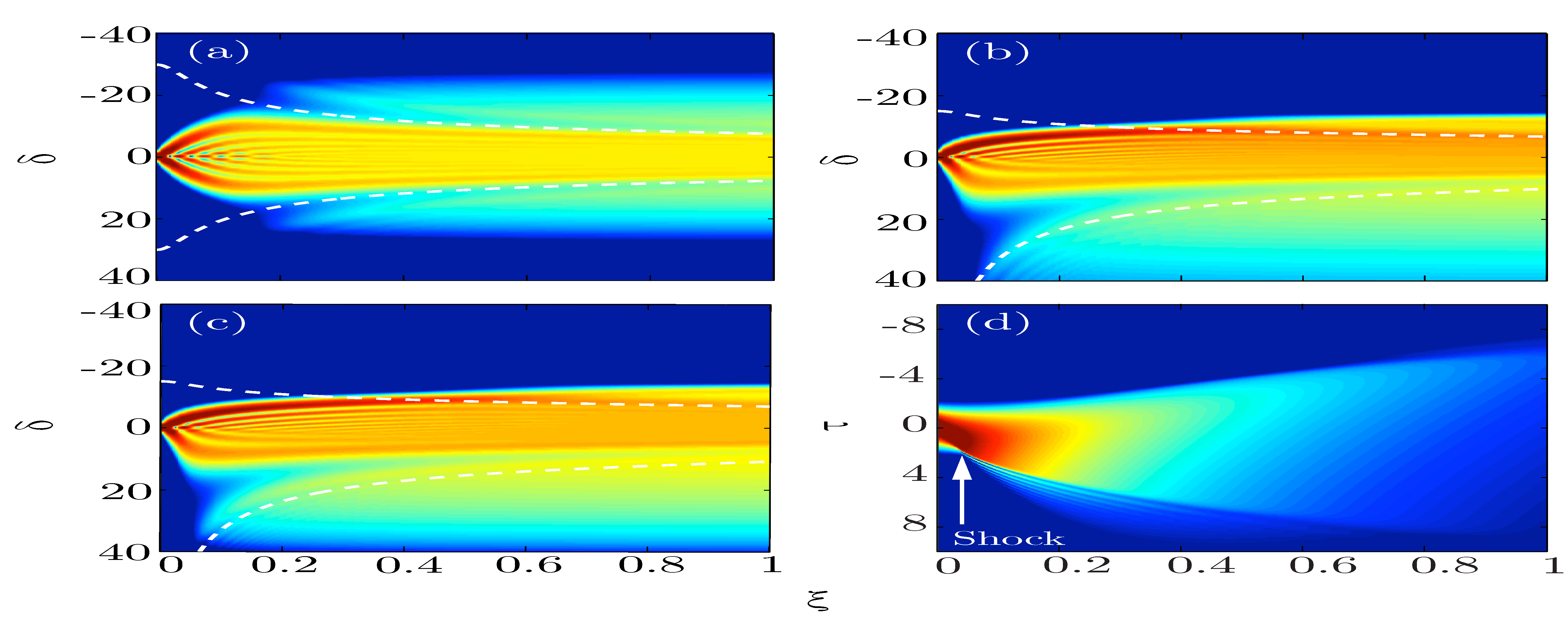}
\caption{(a) Propagation of an input pulse $\psi_{\rm in}=N\sech{\tau}$ in the normal dispersion, with $N=15$, $\alpha_{3}=\mu_{\rm sh}=0$, for one dispersion length, $\xi=1$. (b) Same as (a) but with $\mu_{\rm sh}=0.05$. (c) Same as (a) but with $\tau_{\rm}=0.05$ and $\alpha_{3}=0.005$. (d) Contour plot of the pulse propagation in the time domain, with the same parameters as in (c). Note that the formation of a shock at $\xi\sim0.02$ coincides with the moment of the formation of the two spectral sidebands in (c). Dashed white lines denote the position of the RR calculated by using the phase-matching condition Eq. (\ref{phasematchingadi}). (a-d) Are plotted in logarithmic scale over 2.5 decades. The white arrow in (d) indicates the point of shock formation. %Dashed white line is the position of the zero-GVD point, when present.
\label{fig1}}
\end{figure}

We first note that if the shock term and HOD are neglected, then Eq.~\eqref{fullsolution} leads to:
\eq{onlykerr}{\Delta\omega_{\pm}\simeq\pm\sqrt{2\gamma P/\beta_{2}}}
In view of the existing literature on resonant radiation from solitons, this result is interesting and totally unexpected: one does not need HOD or a zero-GVD frequency point in order to observe resonant radiation, but just $\beta_{2}$ and Kerr are necessary in the normal GVD for \emph{sufficiently high laser pulse powers}. The remaining terms in  Eq.~\eqref{fullsolution} indicate that the presence of the shock term and HOD terms can then strongly unbalance the spectral location of the two resonant radiation peaks, which will shift to one side of the spectrum, making one of them (typically the blue shifted one) very visible both numerically and experimentally, while hiding the other inside the central spectral body of the input pulse.\\
\emph{Numerical simulations ---} We now prove the validity of the phase-matching condition Eq.~(\ref{phasematching}) by means of numerical simulations. First of all we write Eq.~(\ref{phasematching})  in dimensionless units, by using the rescaled dimensionless variables $\xi\equiv z/z_{0}$, $\tau\equiv t/t_{0}$, $\psi\equiv A/\sqrt{P_{0}}$, $z_{0}\equiv t_{0}^{2}/|\beta_{2}|$, $P_{0}\equiv(\gamma z_{0})^{-1}$, $\mu_{\rm sh}\equiv(\omega_{0}t_{0})^{-1}$, $\alpha_{m}\equiv\beta_{m}/[|\beta_{2}|m!t_{0}^{m-2}]$ ($m\ge3$), $\delta\equiv\Delta\omega t_{0}$.
The positions of the resonant radiation frequencies Eqs.~(\ref{fullsolution}-\ref{onlykerr}) in dimensionless units then becomes:
\eq{phasematchingadi}{\sum_{m\ge2}\alpha_{m}\delta^{m}=N^{2}[1+\mu_{\rm sh}\delta]}
\eq{onlykerradi}{\delta_{\pm}\simeq\pm\sqrt{2}N}
\eq{kerrshockadi}{\delta_{\pm}=N^{2}\mu_{\rm sh}\pm\sqrt{(N^{2}\mu_{\rm sh})^{2}+2N^{2}}}
With the above scalings, we then solve Eq. (\ref{NLSE}) in its dimensionless form by means of a standard split-step Fourier scheme, in which the nonlinear part is integrated by means of a fourth order Runge-Kutta algorithm. We use an input pulse $\psi_{\rm in}=N\sech{\tau}$. The results are shown in the panel of Fig. \ref{fig1}.
In Fig. \ref{fig1}(a) we show the spectral evolution of a strong pulse of dimensionless amplitude $N=15$, propagated for $\xi=1$ (i.e. one second-order dispersion length) in the normal dispersion regime, with no higher-order dispersive terms and when neglecting the shock operator, i.e. $\mu_{\rm sh}=0$. After the usual SPM phase, which broadens the spectrum symmetrically, one can notice around $\xi\sim 0.02$, two sidebands that correspond exactly to the resonant radiation bands described by Eq. (\ref{onlykerradi}) (shown by white dashed lines.). In this case, only the normal GVD and the Kerr nonlinearity are present and the two sidebands are perfectly symmetric with respect to $\delta=0$, i.e. the input pulse frequency. This case is conceptually important but unrealistic since the shock term can never be switched off in practice and constant GVD is very difficult if not impossible to achieve in ordinary waveguides or in bulk media.\\
In Fig. \ref{fig1}(b) we still do not introduce any higher-order dispersion term, but we introduce a shock term coefficient $\mu_{\rm sh}=0.05$. Such a value of $\mu_{\rm sh}$ would correspond, in dimensional units to a pulse duration $t_{0}\simeq8.5$ fs, if the pump wavelength is $\lambda_{0}\simeq0.8$ $\mu$m. All other parameters are the same as in Fig. \ref{fig1}(a). One can see that introducing the shock term strongly modifies the position of the sidebands, the positions of which become strongly unbalanced with respect to $\delta=0$. This is also accurately predicted by Eq. (\ref{kerrshockadi}), represented with white dashed lines in the figure. As discussed above, one of the sidebands shifts its position very close to the input pulse frequency, and is  `hidden' inside the spectral body of the self-modulated pulse. The second, blueshifted sideband is much more evident and also possesses a much broader bandwidth than the redshifted band.\\
Fig. \ref{fig1}(c) is the same as Fig. \ref{fig1}(b), but with an additional third-order dispersion term $\alpha_{3}=0.005$ included. It is evident that third-order dispersion does not dramatically affect the position of the sidebands from the previous cases. More complicated cases, when both the full dispersion and the shock term are included, can be studied easily by solving the generally large polynomial equation Eq. (\ref{phasematchingadi}). In Fig. \ref{fig1}(d) we show the space-time contour plot of the propagation relative to Fig. \ref{fig1}(c). One can see that the moment when the sidebands start to be emitted spectrally is exactly when a shock appears in the time domain (indicated with an arrow in in Fig. \ref{fig1}(d)).\\
\emph{Experiments in bulk fused silica glass.}
A remarkable feature of RR is that it can also be accurately described within the framework of a first Born approximation, i.e. it is essentially a linear wave scattering process from a time-dependent (moving at speed $v$) scatterer \cite{kolesikRR1D}. Moreover, this has been shown to be true not only in 1D but also in 2D \cite{kolesikRR2D} and 3D bulk interaction geometries \cite{kolesikRR3D,faccioOE}. In the full 3D geometry, the standard momentum relation for RR in the 1D case becomes a relation for the (propagation direction) z-component of the 3D wavevector, which accounting for the shock term introduced in Eq.~\eqref{fullsolution}, reads as
\begin{equation}
k_z(\omega) = k(\omega_0) + (\omega-\omega_{0})\left[\frac{1}{v}-\frac{\gamma P}{\omega_{0}}\right].
\label{eqn2}
\end{equation}
The full $(\theta,\lambda)$ location of the 3D RR peak is then given by $\theta\sim k_\perp/k=\sqrt{1-(k_z/k)^2}$, where $k=\omega n(\omega)/c$. Experiments may therefore be either carried in 1D fibre geometries or, more conveniently from the perspective of attaining high pulse intensities, also in a 3D bulk geometry. Following this idea, we carried out experiments in bulk fused silica glass. We used an input Bessel pulse in order to achieve the essential conditions required for observing shock-front induced RR: (i) the non-diffracting Bessel peak gives an effective fiber-like propagation regime with sustained peak intensities over long distances, (ii) the  spatial Bessel dynamics lead to an effective suppression of transverse spatial instabilities, in particular self-focusing and  filamentation that typically dominate Gaussian pulse propagation \cite{extreme_bessel}, (iii) high intensities (multi-TW/cm$^2$) are attained due to the suppression of filamentation effects, i.e. suppression of the clamping of the peak intensity to a maximum value that typically occurs with Gaussian pulses \cite{extreme_bessel}.\\
In our experiments, ultrashort optical pulses ($t_{0}\sim100$ fs, central wavelength 785 nm) are shaped into Bessel beams using an axicon of base angle $\theta = 3^{\circ}$ that creates an intense on-axis central peak ($I_{\text{max}}=8$ TW/$cm^2$). This intense peak is normally incident onto a $\sim$1.5 cm thick piece of fused silica. %The beam forms a non-diffracting peak in the Kerr medium and the beam undergoes nonlinear effects along the optical axis, while simultaneously suppressing transverse beam reshaping and self-focussing induced collapse [cite Faccio 2012].  
\\
\begin{figure}
\includegraphics[width = 8.5cm]{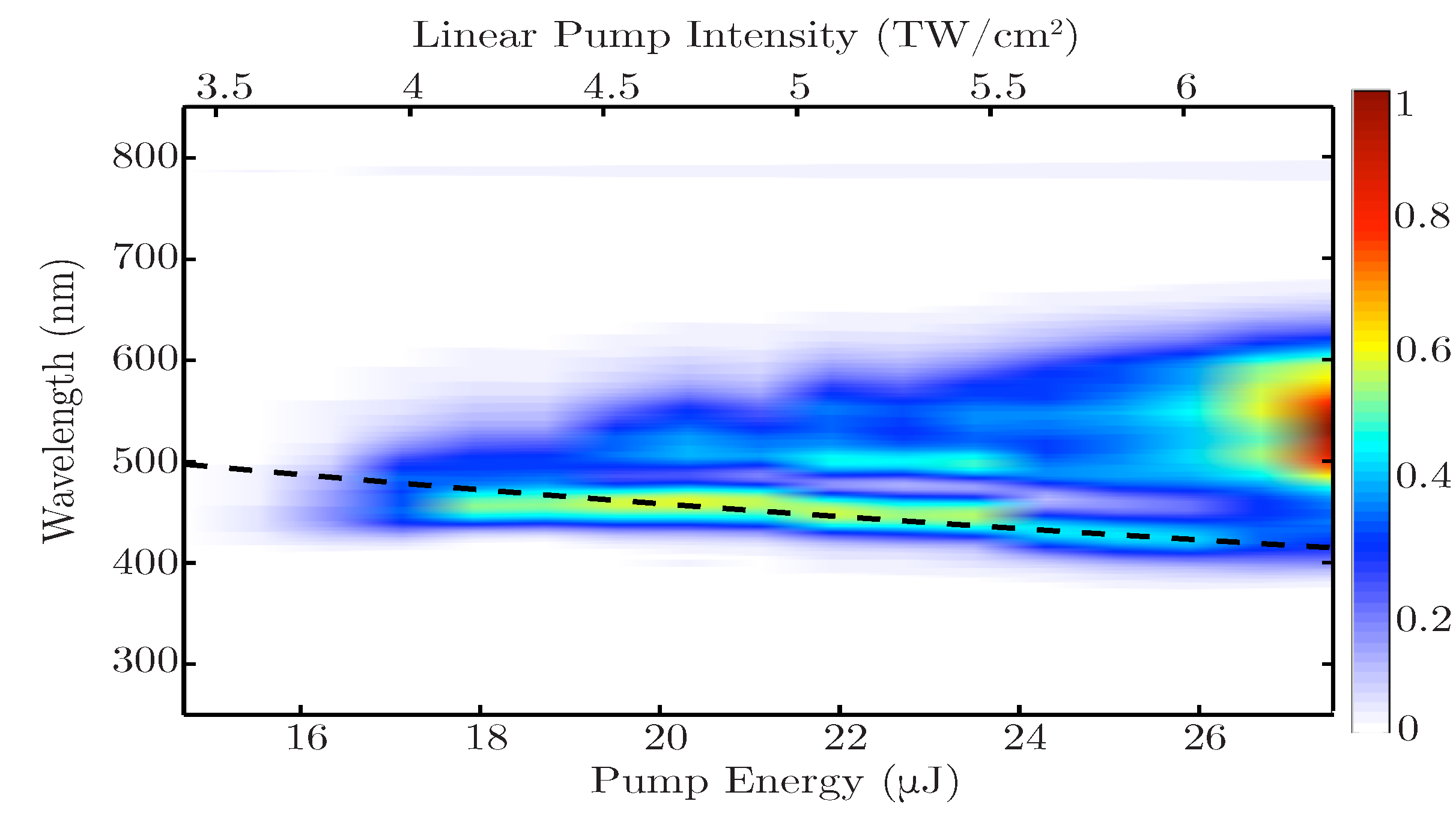}
\caption{Evolution of spectral broadening from a Bessel beam in bulk fused silica as the intensity of the beam is increased. The lower x-axis shows the average energy of the pulses while the upper x-axis estimates the peak intensity of the Bessel beam at the position of the fused silica block. The spectra were measured after a filter that has a sharp long-wavelength cut-off for $\lambda>700$ nm. The input pump pulse spectrum therefore only appears as a very weak feature at 785 nm.  The dashed line shows the fit of Eq.~\ref{eq2}, which fits well over the range of energies used in this study.}
\label{fig2}
\end{figure}
%The transmitted pulses are collected by an imaging telescope and imaged in the far-field onto an imaging spectrometer (Andor Shamrock SR163 equippned with a digital DSLR camera, Sony, modified so as to have increased sensitivity in the near-IR Region) and in the near-field onto a fibre spectrometer (Ocean Optics HD4000). 
In Fig.~\ref{fig2} we show the output spectra measured with a fibre spectrometer (Ocean Optics HD4000), integrated over the width of the beam. A filter that cuts all wavelengths above 700-750 nm is included in order to avoid saturation and damage from the pump beam. The input pump energy is increased ranging from 15 to 27 $\mu$J corresponding, for the Bessel angle and wavelength used here, to intensities ranging from 3.5 to 6.5 TW/cm$^2$ (indicated in the figure). 
At low input energies (15-18 $\mu$J) there is spectral broadening generated by SPM from the input pulse. For higher input energies we observe the formation of a separated peak at a wavelength that blue-shifts with increasing pump energy. At higher input energies ($>26$ $\mu$J), filamentation dynamics finally take over and supercontinuum generation washes out the separated peak around 400 nm. We interpret the isolated blue spectral peak (for energies $<26$ $\mu$J) as evidence for shock-induced resonant radiation. Indeed, we fit this peak by using Eq.~\eqref{fullsolution}, which provides an equation for the wavelength of the RR: 
\begin{eqnarray}
\lambda_{\text{RR}} &=& \frac{\lambda_0}{1+x+\sqrt{x(2+x)}} \nonumber\\
\textrm{where} \hspace{0.2cm}x &=& \frac{n_2 I k_0 \lambda_0^2}{2\pi^2c^2|\beta_2|}\text{.}
\label{eq2}
\end{eqnarray}
where $n_2$ is the nonlinear refractive index of the fused silica. %This equation is dependent only on the peak Bessel intensity, % ($I = \gamma P / n_2 k_0$),
We underline that when fitting with this equation, there are no free parameters, yet,  as shown by the dashed black line in Fig.~\ref{fig2}, this relation provides an excellent fit to the experimental data.  %It is interesting to note that the pump intensity (estimated within the bulk silica, which cannot be directly measured) is an order of magnitude greater than that expected for a Gaussian pulse for an equivalent focal length lens. This may explain why this kind of RR has yet to be observed experimentally.
 In the supplementary material we compare measurements for the Bessel pulse (shown in Fig.~\ref{fig2}) with an input Gaussian pulse. The Gaussian pulse shows evident filamentation dynamics and supercontinuum generation for input peak intensities that are more than one order of magnitude lower than the observed Bessel beam filamentation threshold. Therefore, with a Gaussian beam the shock-induced RR is either simply not excited (due to the lower peak intensities) or may still be present but in any case is covered by the filament supercontinuum and is not distinguishable as a separate peak. Conversely, with Bessel beam filamentation, supercontinuum occurs only at much higher intensities. %thus allowing the generation of a sufficiently blue-shifted RR peak that is not covered by the supercontinuum.\\
%It is worth noting that the shock-induced RR is quite different in nature from the negative-frequency resonant radiation created in the presence of very steep shock fronts that we have reported recently \cite{NRR,SR}. Indeed, the NRR for these experiments is predicted to lie at significantly shorter wavelengths in the UV region and is not visible in these experiments (see supplementary information for more details). \\ 
%
We note that our experiments in bulk media allow to measure the full space-time spectra in $(\theta,\lambda)$ coordinates \cite{kolesikX,FaccioX1,faccioOE}. 
% We also neglect the nonlinear correction, $- k_{NL}(\omega_0)$ \textit{why?}. 
%
\begin{figure}[t!]
\includegraphics[width = 8.5cm]{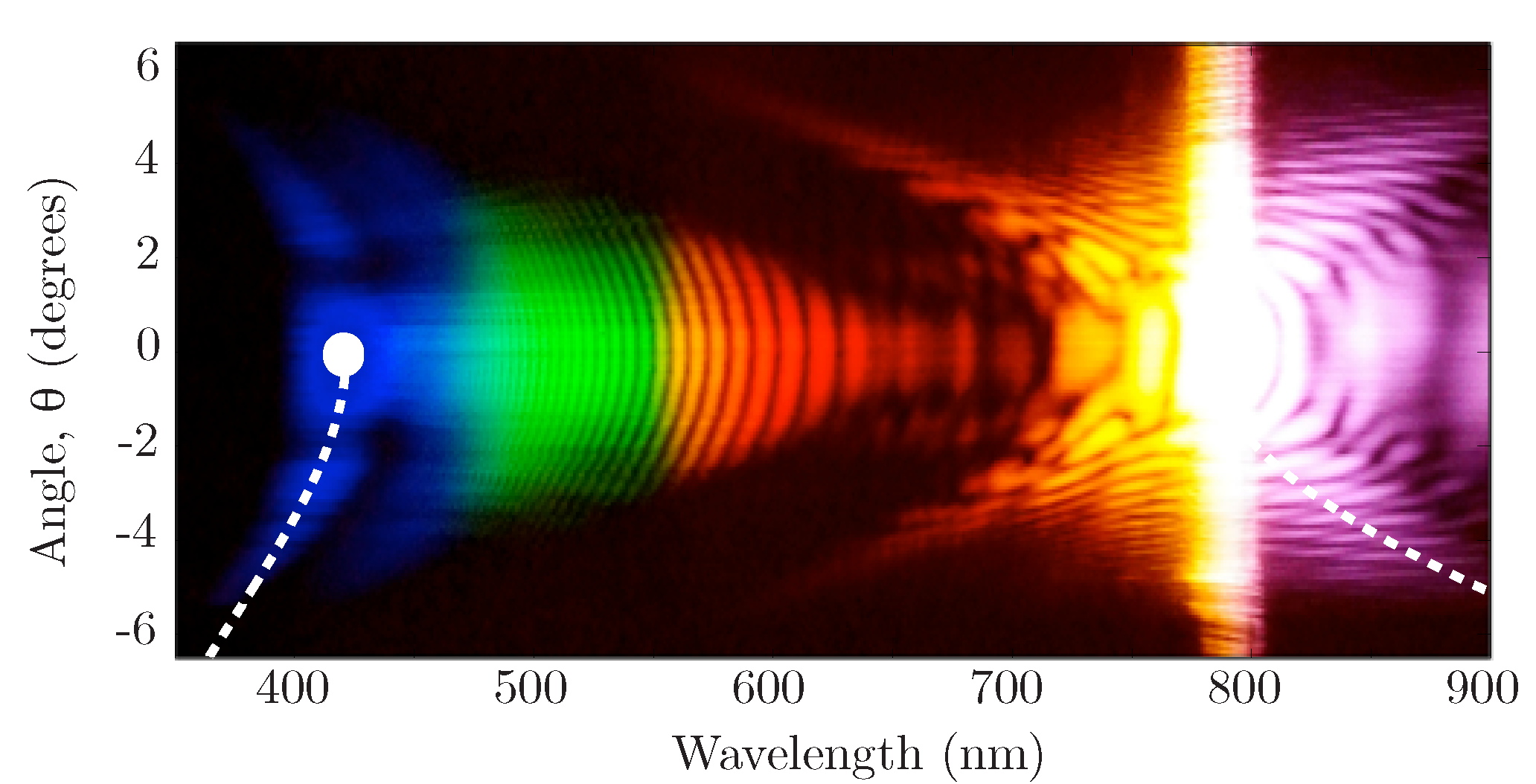}
\caption{Measured spectrum at a pump intensities of $\sim$21 $\mu$J following formation of the shock-front which induces the RR described by Eq.~\eqref{fullsolution}. Dashed white line - fit from Eq.~\eqref{eqn2}. Solid white circle - 1D solution from Eq.~\eqref{fullsolution}.}
\label{fig3}
\end{figure}
 In Fig.~\ref{fig3} we show the far-field $(\theta,\lambda)$ spectrum  for a pump energy of $\sim$21 $\mu$J. This spectrum is measured using a home-built imaging spectrometer and a modified digital camera. The latter allows to measure spectra without the need for any additional filters to cut the pump beam, thus the whole frequency spectrum is now clearly visible. Figure~\ref{fig3} shows a clear, intense blue-shifted peak close to 400 nm wavelength that is separated from the rest of the spectrum and whose position and shape is well reproduced  by Eq.~\eqref{eq2} (dashed white line) where the peak intensity is  taken as the intensity at the sample input facet  (see supplementary information). The full white circle in the figure shows the calculated wavelength for shock-induced resonant radiation as predicted by the $\Delta\omega_+$ solution to the 1D relation, Eq.~\eqref{fullsolution}. This solution naturally coincides with the full 3D fit for $\theta=0$ deg.
The full far-field evolution (for increasing input energy) of the spectral broadening is shown in the  supplementary material. \\
\emph{Conclusions ---} We have demonstrated theoretically and experimentally  a novel form of resonant radiation that is the result solely of the high laser pulse intensity and is highly visible in the presence of a shock front in the form of a blue-shifted spectral peak. This form of wave-breaking finds a quantitative description in terms of a generalised linear-wave resonant scattering process that occurs in normal GVD and even in the absence of a zero GVD point. The implications are that RR is intrinsically linked to wave breaking and therefore also to self-steepening and shock front formation processes. Therefore, RR is not only a soliton effect, as originally thought and widely presented in the literature. Rather, it is associated to the mechanism by which a wave packet stabilises itself by resonantly shedding light to any other allowed mode on the dispersion curve. With negative GVD we would observe the traditional soliton RR. While in normal GVD, shock fronts will also form and will stabilise themselves (i.e. the wave ``breaks") by RR emission. The ubiquity of similar shock front dynamics in other systems, e.g. gravity waves in water or in plasma waves, points to similar wave scattering processes that should also appear in the same form as described here. \\

D.F. acknowledges  financial support from the Engineering and Physical Sciences Research Council EPSRC, Grant EP/J00443X/1 and from the European Research Council under the European Union's Seventh Framework Programme (FP/2007-2013) / ERC Grant Agreement n. 306559.

%The most surprising result is that this radiation can be emitted even in absence of a zero-GVD point. The presence of a shock front, that will usually accompany pulses with high intensities, is shown to modify the spectra location of the resonant instability that therefore will typically appear as an isolated blue shifted peak in the (normal GVD region of the) spectrum. 
%The ability to control such radiation can lead to novel ways to implement and control supercontinuum generation in the normal GVD regime, which is modulationally stable. A modulationally stable background is important at a classical level as it allows generation of supercontinua with higher coherence. Likewise, at the quantum level, we expect the recently discovered vacuum squeezed states associated with RR to be robust against decoherence associated to modulation instability \cite{biancalana_squeezing}.

\end{document}